\documentclass[twocolumn,showpacs,preprintnumbers,amsmath,amssymb]{revtex4}

\usepackage{graphicx}
\usepackage{dcolumn}
\usepackage{bm}

\begin{document}

\preprint{APS/123-QED}

\title{Long radial coherence of electron temperature fluctuations in non-local transport in HL-2A plasmas}

\author{Zhongbing Shi}
\email{shizb@swip.ac.cn}
\affiliation{Southwestern Institute of Physics, Post Office Box 432, Chengdu 610041, China}
\author{ Kairui Fang}
\affiliation{Southwestern Institute of Physics, Post Office Box 432, Chengdu 610041, China}

\author{Jingchun Li, Zhaoyang Lu}

\affiliation{Department of Earth and Space Sciences, Southern University of Science and Technology, 518055 Shenzhen, Guangdong, People's Republic of China}%

\author{Xiaolan Zou}
\affiliation{CEA, IRFM, F-13108 Saint-Paul-lez-Durance, France}%

\author{Jie Wen, Zhanhui Wang, Xuantong Ding, Wei Chen, Zengchen Yang, Min Jiang Xiaoquan Ji, Ruihai Tong, Yonggao Li, Peiwang Shi, Wulyv Zhong, and Min Xu}
\affiliation{Southwestern Institute of Physics, Post Office Box 432, Chengdu 610041, China}

\date{\today}

\begin{abstract}
The dynamics of long-wavelength ($k_\theta<1.4 \mathrm{\ cm^{-1}}$), broadband (20-200 kHz) electron temperature fluctuations ($\tilde T_e/T_e$) of plasmas in gas-puff experiments were observed for the first time in HL-2A tokamak. In a relative low density ($n_e(0) \simeq 0.91 \sim 1.20 \times10^{19}/m^3$) scenario, after gas-puffing the core temperature increases and the edge temperature drops. On the contrary, temperature fluctuation drops at the core and increases at the edge. Analyses show the non-local emergence is accompanied with a long radial coherent length of turbulent fluctuations. While in a higher density ($n_e(0) \simeq 1.83 \sim 2.02 \times10^{19}/m^3$) scenario, the phenomena were not observed. Furthermore, compelling evidence indicates that $\textbf{E} \times \textbf{B}$ shear serves as a substantial contributor to this extensive radial interaction. This finding offers a direct explanatory link to the intriguing core-heating phenomenon witnessed within the realm of non-local transport.
\end{abstract}

%
%
%
%
%

\maketitle

\section{\label{sec:level1}Introduction}

One of the perplexing phenomena observed and extensively studied worldwide is non-local transport, which has been investigated in various tokamaks (TFTR\cite{kissick1994transient},TEXT\cite{gentle1995strong}/J-TEXT\cite{Shi_2018}, RTP\cite{galli1999non,mantica1999nonlocal}, HL-2A\cite{hong2007observation}, Alcator C-mod\cite{rice2013non}, KSTAR\cite{shi2017intrinsic}, EAST\cite{liu2019dynamics}, Tore Supra\cite{zou2000edge} , JET\cite{goriniperipheral}, AUG\cite{ryter2000propagation}, DIII-D\cite{rodriguez2019predict}) and stellerators(W7-AS\cite{stroth1996}, LHD\cite{inagaki2006abrupt}). These experiments have consistently revealed a peculiar long-range transient temperature response triggered by pulsed interventions. These observations suggest that disturbances at the plasma's edge can perturb the deeper plasma equilibrium, challenging the conventional understanding of transport governed by local parameters and the gradient-flux relation\cite{ida2015towards, Inagaki_2013}.

In recent years, turbulence theories focusing on multiple-scaled nonlinear and statistical dynamics have gained traction in addressing this enigma. These include turbulence spreading models\cite{hahm2004turbulence,gurcan2005dynamics,lin2004turbulence,wang2011turb,hariri2016cold} and the concept of self-organized criticality (SOC)\cite{diamond1995dynamics,kubota1997avalanche,wang2011turb,chen2016dynamics}. Moreover, several studies propose that the transition between trapped electron modes (TEMs) and ion temperature gradient modes (ITGs) may play a role in the critical nature of non-local effects\cite{gao2014non}. Evidence has emerged indicating the coexistence of rotation reversal, shear, and heat pinch phenomena\cite{rice2013non,shi2017intrinsic,ida2015towards,mantica2000heat}. Notably, in the past year, a new form of non-locality in the ion temperature channel was discovered in J-TEXT\cite{Shi2020Observation}.

Despite more than two decades since its discovery, the mechanisms and theoretical models behind non-locality remain enduring enigmas in the field. In recent years, a particular perspective has emerged, challenging the traditional understanding of non-local transport. This perspective, supported by simulations, posits that the injection of a cold pulse may not actually induce genuine non-local transport\cite{rodriguez2018explaining, Angioni_2019}. Instead, it suggests that non-locality can be explained as an outcome of the transient changes in local density gradients, with distinct mechanisms governing such behavior in plasmas dominated by TEMs versus ITGs\cite{li2019,li2022,li2023}.

Undoubtedly, local drift-wave turbulent transport mechanisms stand as widely accepted explanations for anomalous transport phenomena in tokamaks. Gyrokinetic and gyrofluid theories have been extensively validated and serve as robust frameworks for understanding these processes\cite{li2023}. However, intriguingly, observations from experiments like EAST have revealed instances where density fluctuations transiently increase in the core before any local alterations in density and temperature occur\cite{liu2019dynamics}. Such findings hint at the possibility of a turbulence spreading model\cite{hariri2016cold} at play. The question that looms large is how these turbulent fluctuations can evolve more rapidly than local profiles or gradients, a puzzle that demands further elucidation.


Hence, non-locality, to a certain extent, remains an enigma within the domain of heat transport. This perplexity arises from the challenges in deciphering thermal fluctuations within the turbulence. These fluctuations are characterized by their low amplitude within the core region, their propensity for long-distance effects, and the formidable obstacles posed by localized measurements. Notably, the turbulence and its electron thermal fluctuation (referred to as $\tilde T_e/T_e$) present formidable detection and measurement challenges. This is exacerbated by the fact that the inherent thermal noise level is quite substantial, exceeding 4$\%$ of the temperature level, in stark contrast to the 1$\%$ threshold for $\tilde T_e/T_e$.

Thankfully, a ray of hope has emerged with the successful adaptation of correlation techniques initially developed for astronomical detection instruments into the realm of plasma diagnostics. A notable development on this front occurred last year with the deployment of a correlation electron cyclotron emission system (CECE) in the HL-2A tokamak. This advancement has played a pivotal role in shedding light on the profound physics underlying the non-local phenomenon. In this letter, we present a detailed analysis of electron temperature fluctuations, unveiling several groundbreaking findings for the first time.

\begin{figure}[!ht]
	\centering
	\includegraphics[scale=0.40]{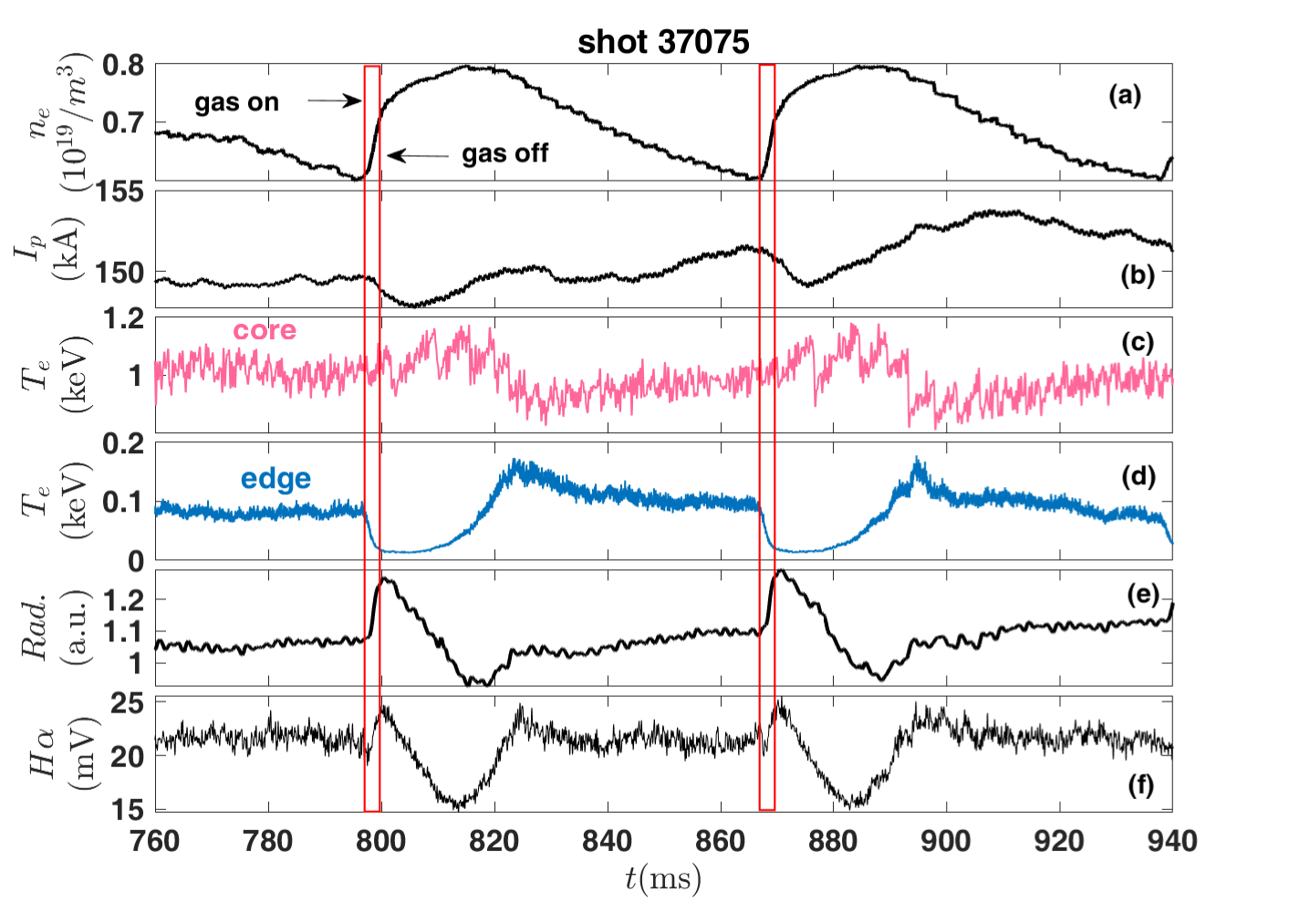}
	\caption{\label{fig:fig1}Raw data in a non-local experiment (shot 37075). Time evolutions of (a)electron density from far infrared interferometer(FIR). (b)Plasma current. Electron temperature from ECE at (c) normalized radii $\rho \sim 0.10$ and (d) $\rho \sim 0.80$. (e) Power of plasma radiation from bolometer and (f)Raw signal of $H_{\alpha}$ radiation diagnostic. The perpendicular red lines represent the injection on and off, the same below.}
\end{figure}

\section{Experimental Setup}

This study is based on data obtained from the HL-2A tokamak, a medium-sized device with a major radius (R) of 1.65 meters and a minor radius (a) of 0.4 meters \cite{LIU2003147}. The tokamak is equipped with a toroidal magnetic field ranging from 1.2 to 2.7 Tesla and is capable of generating plasma currents of up to 480 kiloamperes. The versatility of HL-2A allows for the investigation of non-local transport effects through various experimental techniques, including modulated electron cyclotron heating (ECH), supersonic molecular beam injection (SMBI), pellet injection, gas puffing, fast current ramps, and laser blow-off \cite{sun2008features}. In this particular study, we focus on non-local phenomena induced by gas puffing (GPs) in ohmic-heating plasmas characterized by relatively low electron densities (approximately $0.91-1.20 \times 10^{19}/m^3$). Key diagnostic tools employed include a multi-channel Electron Cyclotron Emission (ECE) system, which shares a common sightline and horn antenna with the CECE system, and a Doppler BackScattering (DBS) reflectometer located at a toroidal angle of 22.5 degrees relative to the ECE window \cite{zhongbing2018progress}. The DBS instrument provides valuable information about plasma rotation, density fluctuations, and radial electric fields.


\begin{figure}[!htbp]
	\centering
	\includegraphics[scale=0.47]{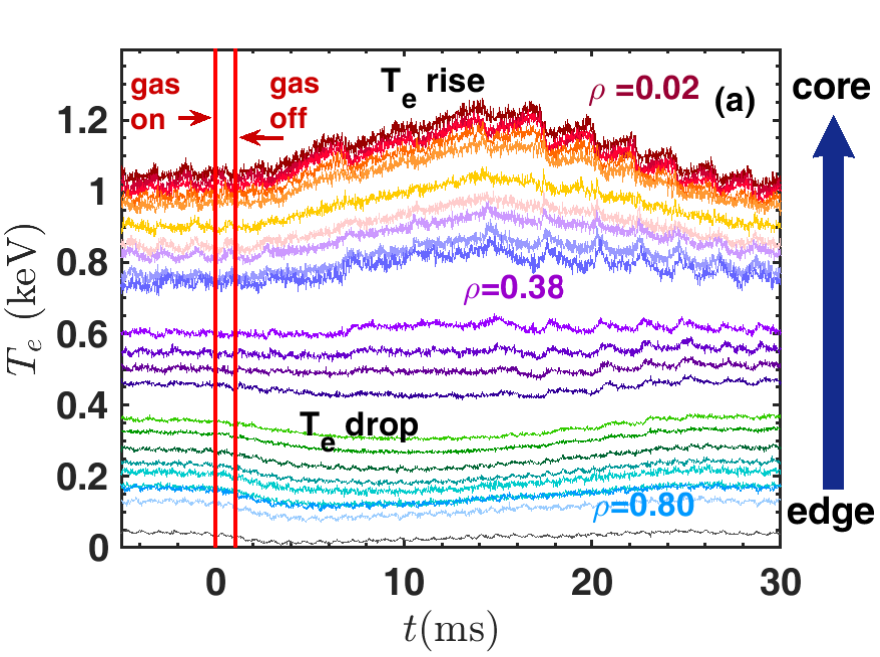}
	\includegraphics[scale=0.47]{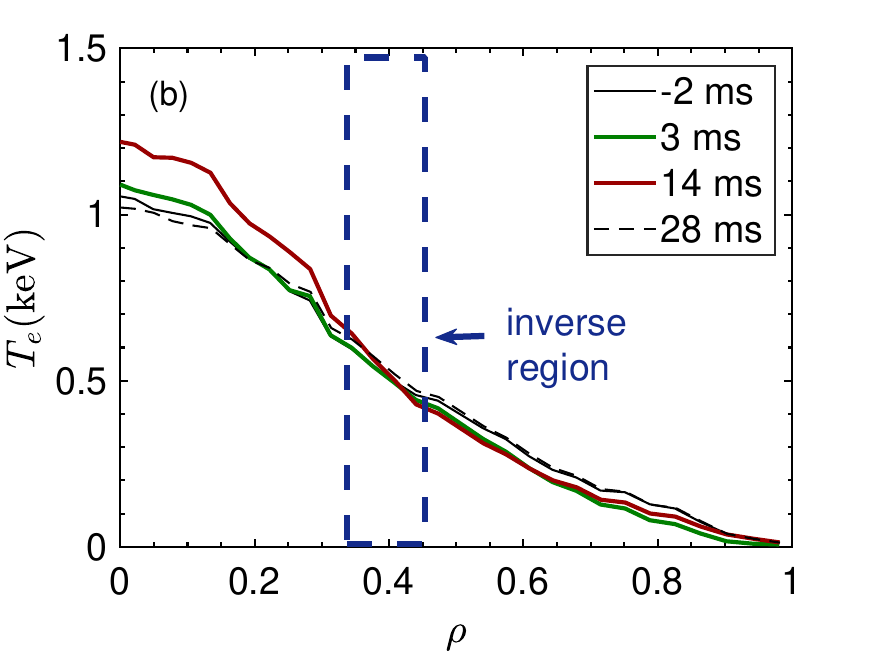}
	\includegraphics[scale=0.47]{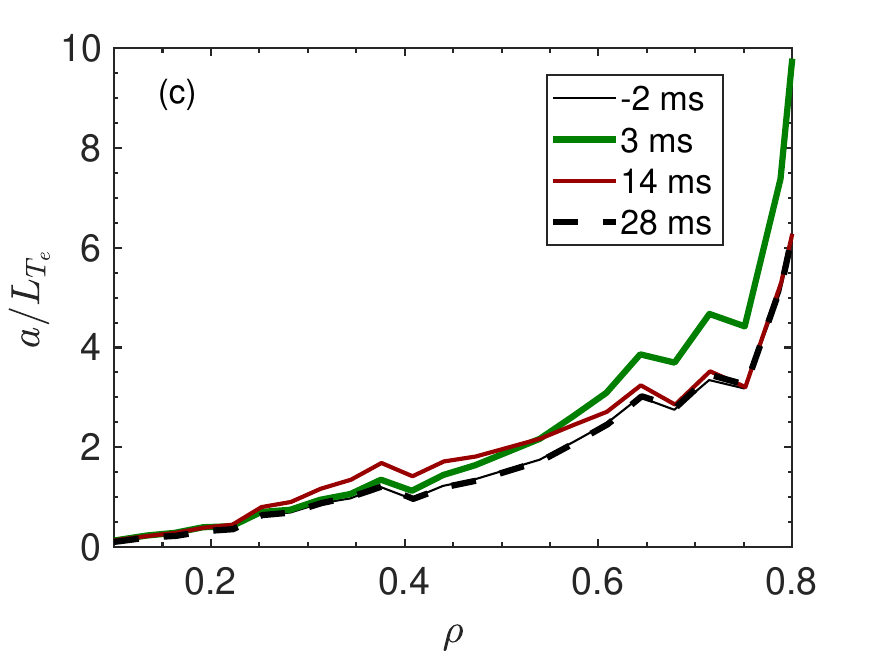}
	\caption{\label{fig2} (a) Electron temperature from plasma edge to the core. The gas is fed at 0 ms. Profiles of (b) electron temperature and (c) normalized $T_e$ gradient scale length at 2 ms before, 3 ms after , 14 ms after ($T_e(0)$ peaking) and 28 ms after (recovery).}
\end{figure}

\section{Experimental results}

\subsection{General non-local phenomenon in GP experiment}
Figure 1 illustrates various parameters observed during a typical discharge involving multiple Gas Puffs (GPs). In this experimental shot, the plasma is subjected to ohmic heating, featuring a toroidal magnetic field configuration of 1.34 T and a plasma current of approximately 150-153 kA. Each gas puff lasts for about 1-2 ms, which is similar in duration to Supersonic Molecular Beam Injections (SMBIs) but with lower pulse intensity. The on/off timing of the GPs is indicated by the front and rear red lines perpendicular to the time axis. It's crucial to note that the time intervals between these pulses are sufficiently long to ensure the independence of each non-local transport effect.

Additionally, the plasma density (Figure 1(a)) is measured using a far-infrared interferometer (FIR) \cite{li2017new}. It's evident that the line-averaged density increases immediately after the GPs and then gradually decreases. Simultaneously, the inner core electron temperature exhibits a bulge, followed by a recovery over a duration of approximately 30 ms, while the edge amplitudes undergo opposite changes. The evolution of plasma current, radiation power, and $H_\alpha$ signals are also displayed. 

The foundational investigation commences with equilibrium parameters because local collisionality and gradients in the background are likely to influence the turbulent state, subsequently affecting the measurements. Figure 2(a) presents the time traces of electron temperature ($T_e$) signals at different radial positions. It's important to emphasize that the puff time is normalized to zero with the onset of gas injection, facilitating a clearer time series for signal comparison. The heat diffusivity induced by sawtooth oscillations ranges from 4-6 $m^2/s$ at $\rho \simeq 0.5$, with no discernible variation before and after the occurrence of non-local phenomena. The puffs in the experiment are deployed beyond $\rho > 0.7$, resulting in immediate changes in edge profiles. Figure 2(b) and 2(c) display profiles of electron temperature ($T_e$) and the normalized gradient scale length $a/L_{T_e}$ before, during, and after the non-local effect. Negative time values represent the period before puffing when parameters serve as references. In Figure 2(b), we observe a rapid change in $T_e$ occurring first outside the inverse radius, peaking with a delay of approximately 2-3 ms. Subsequently, the core region of $T_e$ begins to rise, with a maximum increment of about $18\%$ (approximately 200 eV) observed in the central core at around 14 ms. Regarding the normalized gradient scale length $a/L_{T_e}$ presented in Figure 2(c), local $a/L_{T_e}$ variations are consistent with local $T_e$ over time. In contrast, $a/L_{T_e}$ displays a global increase following the gas puff.

\begin{figure*}[ht]
	\centering
	\includegraphics[scale=0.6]{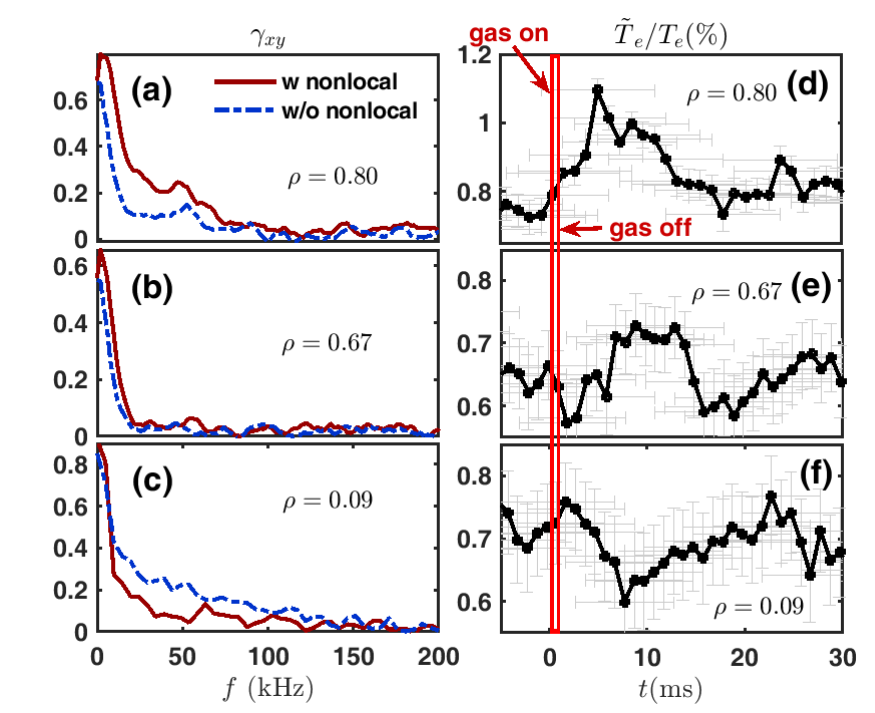}
	\caption{\label{fig3}(a-c)Cross coherencies caculated during the nonlocal phases (w nonlocal) and the steady state phases (w/o nonlocal) at different radii and(d-f)Time sequences with $\tilde T_e/T_e$ (20-200kHz).}
\end{figure*}

\subsection{Non-locality in electron temperature fluctuations}
Figure \ref{fig3}(a-c) illustrates the coherencies, while figure \ref{fig3}(d-f) displays normalized time-domain fluctuations ($\tilde T_e/T_e(t)$) integrated within the 20-200 kHz frequency range. Evidently, following the gas puff, there is a rapid and significant increase in the amplitude of relative electron temperature fluctuations, primarily observed at the peripheral radius around $\rho \simeq 0.80$. This observation is substantiated by both the coherency graphs and the $\tilde T_e/T_e(t)$ plots. Moving closer to the location of the $T_e$ inverse radius, there is a diminished distinction between the two phases. Conversely, at the edge, during the non-local phase, $\tilde T_e/T_e(t)$ decreases in the central core. Reflecting on the previous $T_e$ response, the behavior of these fluctuations aligns with the $T_e$ increment/decrement, as a greater/lower fluctuation amplitude corresponds to stronger/weaker turbulence and more/less heat transport (heat loss), assuming other conditions remain constant. Notably, the CECE-measured fluctuations have been previously linked to $\nabla T_e$ driving in ohmic heated L-mode plasma, a concept widely accepted in drift-wave turbulence theories\cite{fang2019eight}.

\begin{figure}[ht]
	\centering
	\includegraphics[scale=0.40]{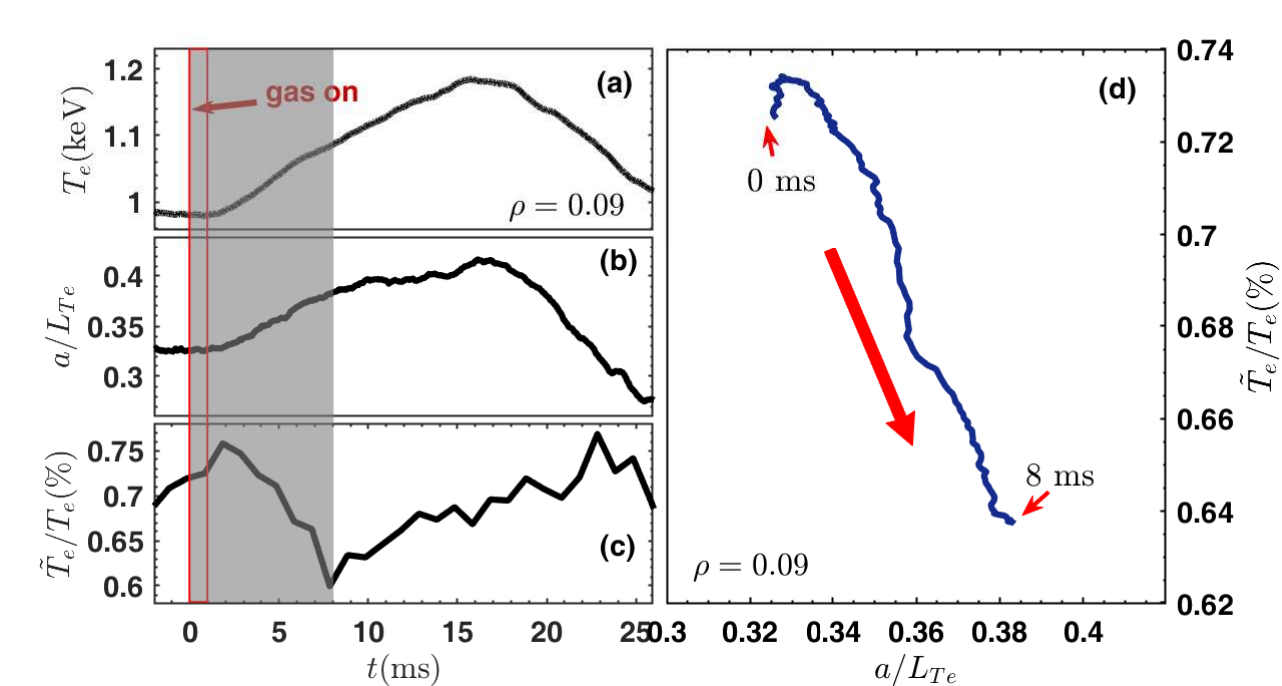}
		\caption{\label{fig4}Evolvement of local ($\rho=0.09$) (a) electron temperature (b) normalized gradient length of $a/L_{T_e}$ and (c) relative fluctuations of $\tilde{T}_e/T_e$. (d) Relation between $a/L_{T_e}$ and $\tilde{T}_e/T_e$. Several core data points imply fluctuations are not mainly driven by the temperature gradients.}
\end{figure}
\begin{figure}[!htbp]
	\centering
	\includegraphics[scale=0.48]{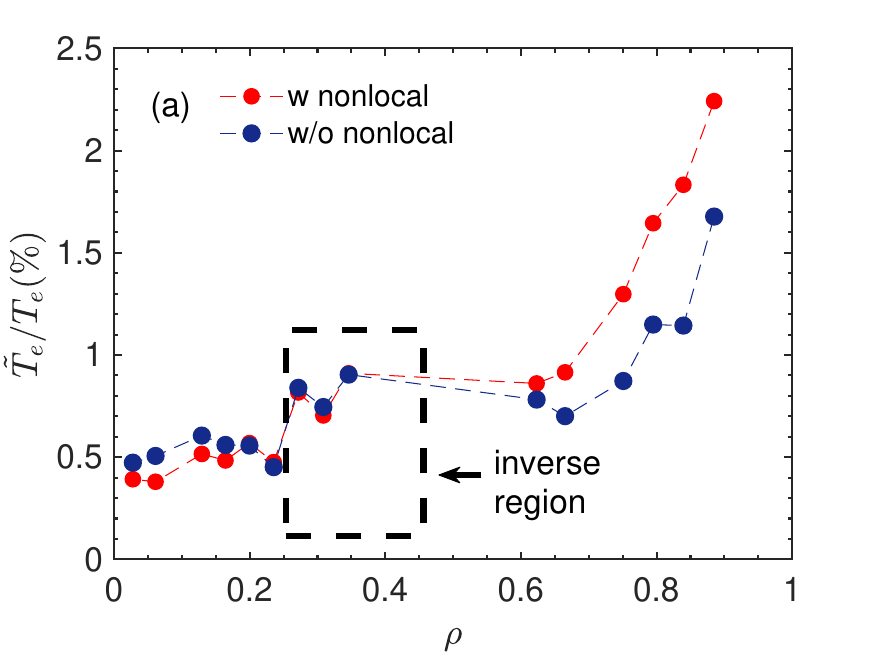}
	\includegraphics[scale=0.48]{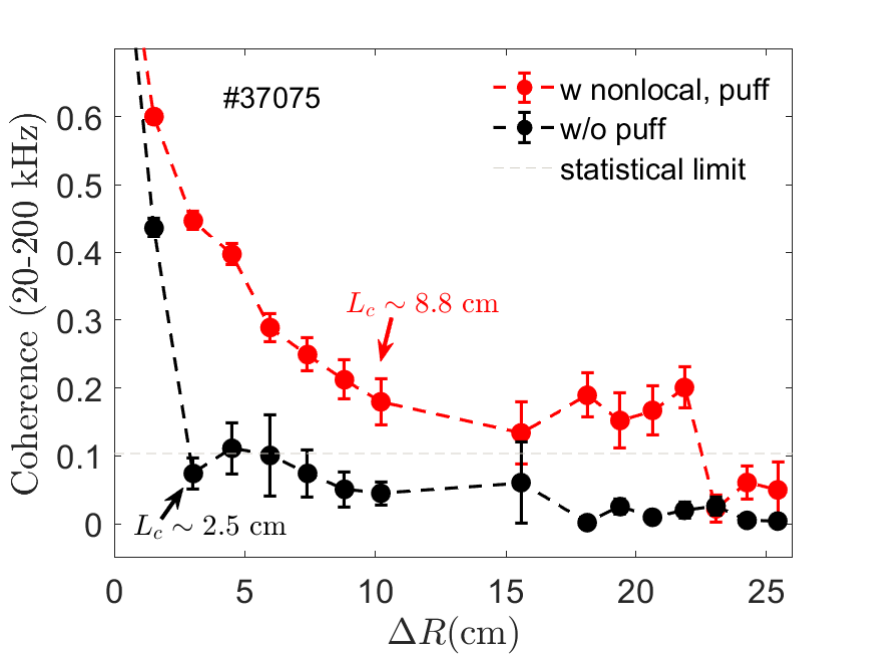}
	\includegraphics[scale=0.48]{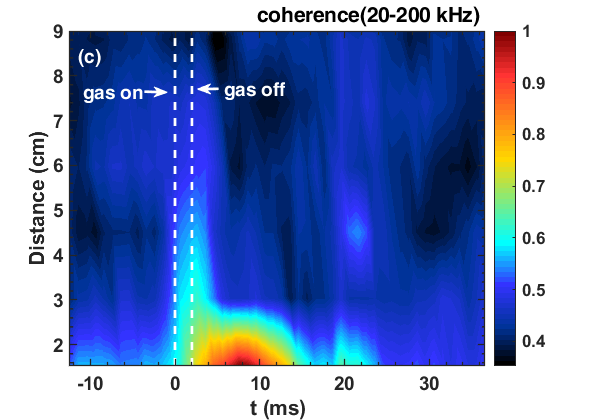}
	\caption{\label{fig5}(a)Relative electron temperature fluctuation profile of $\tilde{T}_e/T_e(\%)$ (20-200kHz) . (b) Radial coherent length with reference position at  $\rho=0.80$ and (c) spatiotemporal structure of coherent length in the low density scenario.  }
\end{figure}

\begin{figure}[!htbp]
	\centering
	\includegraphics[scale=0.48]{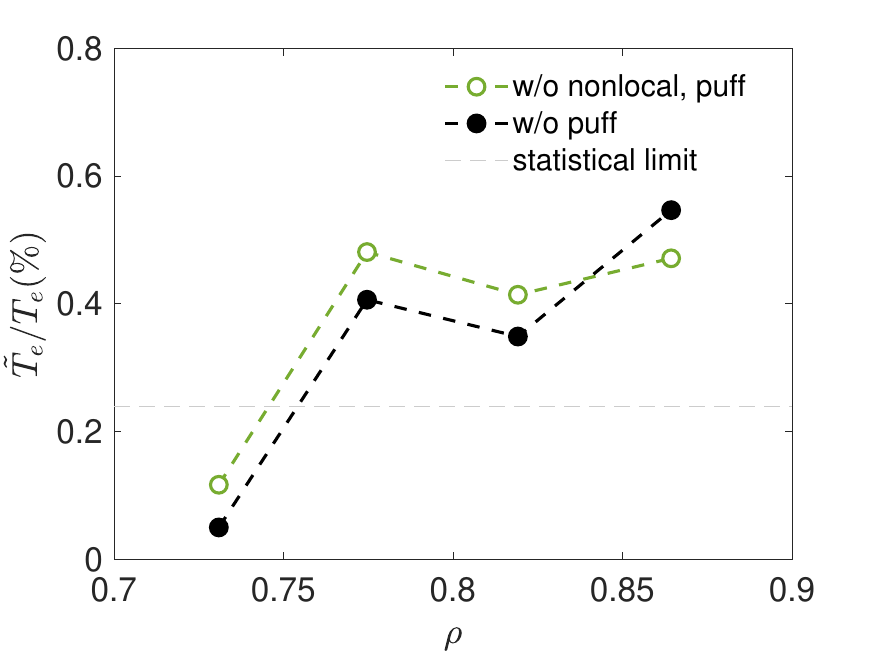}
	\includegraphics[scale=0.48]{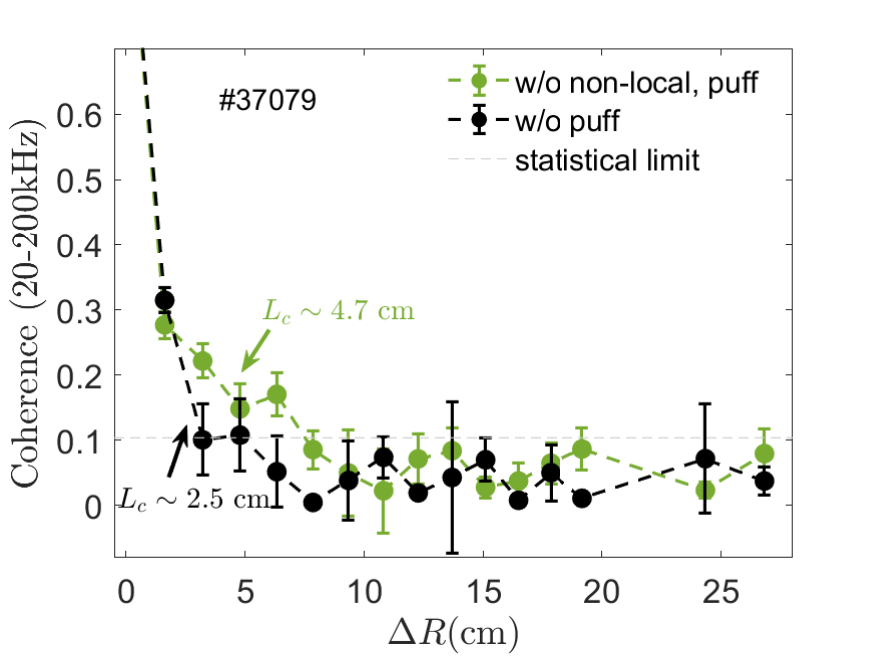}
	\caption{\label{fig7}(a)Profile of $\tilde{T}_e/T_e$ .(b) Radial coherent length with reference of $\rho=0.86$ in the high density ($n_e(0) \simeq 1.83 \sim 2.02 \times10^{19}/m^3$)  scenario. There is no obvious difference between phase before and after the gas puff.}
\end{figure}

This study also undertakes a similar comparison between $a/L_{T_e}$ and $\tilde T_e/T_e$. One of the outcomes, at $\rho \sim0.09$, is presented in figure \ref{fig4}. Interestingly, unlike the typical gradient-flux relation, fluctuations in this scenario exhibit a decrease in amplitude alongside the local gradient of $a/L_{T_e}$, with the latter remaining relatively unchanged (slightly increasing). This finding suggests that fluctuations in the region of $\rho<0.30$ are not primarily determined by local $a/L_{T_e}$; in other words, the steady-state gradient driving is not applicable in this context. For electrostatic turbulence, another possibility considered is the role of $\nabla n_e$ driving. However, validating this mechanism presents challenges, primarily due to limited means of locally measuring core density. While the interferometer with tomographic inversion can reconstruct the density profile, it may introduce distortion in the signal evolution in the core due to edge data interference. Conversely, in the peripheral region, the local gradient and $\tilde T_e/T_e$ can be associated since the relationship between figure \ref{fig2}(c) and \ref{fig3}(d) demonstrates a positive correlation. Notably, no significant changes were observed in coherency and fluctuation evolution around the $T_e$ inverse radius.

Using multiple radial positions, the profile of $\tilde T_e/T_e$ can be constructed, as depicted in Figure \ref{fig5}(a). In light of the preceding analyses, it becomes apparent that the $\tilde T_e/T_e$ profile also exhibits an inverse radius when computed with and without the non-local phases. Furthermore, additional radial cross coherency analyses were conducted, focusing on specific channels in relation to others, as illustrated in Figure \ref{fig5}(b). Specifically, the channel at $\rho=0.80$ was selected as a reference point. The radial coherence length ($L_c$) was theoretically defined as the radial span over which the integrated coherencies or fluctuations diminish to $1/e$ of their amplitude.

Following the gas puffing, the radial coherence length experiences a significant increase during the non-local phases, as evident in Figure \ref{fig5}(b). During the steady phase, the observed radial coherence extends approximately 2.5 cm, but this length escalates to about 8.8 cm during the non-local phase. Given the selected channel at $\rho=0.80$, this result suggests that the long radial influence penetrates to approximately $\rho \sim 0.55$, precisely within the vicinity of the $T_e$ inverse region. Furthermore, the observed radial coherence effect persists, manifesting as a non-Gaussian tail extending approximately 22 cm, signifying profound and deeper transient interactions. This phenomenon also implies that long radial coherent turbulences are established immediately after the gas puffs [Figure \ref{fig5}(c)].

In conjunction with the response time illustrated in Figure \ref{fig3}(d-f), the alterations in coherent length are consistent with fluctuations in amplitude. These observations of long-distance coherence or correlation likely constitute a mechanism for radial transient transport. Furthermore, $L_c$ is determined by integrating fluctuations within the 20-200 kHz frequency range to mitigate the influence of low-frequency MHD interference. This represents a significant advancement compared to previously reported macroscopic structures in some instances.

It is noteworthy that the measurement principle of ECE involves utilizing two adjacent channels with very close frequencies for correlation analysis to eliminate the influence of noise. As illustrated in Figure 1 (see ref. [35]), the bandwidth of the adjacent channels is 150 MHz, corresponding to a spatial width of approximately 1 cm. Due to radiation broadening, there is some overlap between measurements from these two channels. When the two channels are measured independently, turbulence exhibits spatial correlation, while noise lacks correlation, thus achieving the objective of noise reduction. Moreover, in the high-frequency range, the correlation is relatively weak. Although the correlation is weak around 100-200 kHz at $\rho \sim 0.8$, it remains evident when comparing cases with and without non-local phenomena. Additionally, to eliminate the influence near zero frequency, signals in the 0-20 kHz range were not considered. Therefore, a range of 20-200 kHz was employed in this study. In fact, the choice of frequency range has no impact on the results presented in Figure 3.

\subsection{A comparison between low- and high-density scenario}
Before delving into comparative scenarios, it is imperative to juxtapose the performance of $\tilde T_e/T_e$ with density fluctuations derived from DBS data. Within the DBS system, the meso-scale portion (20-100 kHz) of the spectrum demonstrates a degree of consistency with $\tilde T_e/T_e$. However, it is worth noting that due to the distinct wavenumbers detected (with $k_\perp$ falling in the range of 4-10 $\mathrm{cm^{-1}}$ by the DBS and $k_\theta$ remaining below 1.4 $\mathrm{cm^{-1}}$) via the CECE system, higher-frequency ($\le$100 kHz) density fluctuations exhibit dissimilar behavior. This observation aligns with previous reports from HL-2A\cite{shi2018roles,ida2015towards}.

It's important to highlight that cross-correlation analysis between $\tilde n_e$ and $\tilde T_e$ was not conducted in this study. Nonetheless, an alternative and valuable physical parameter can be derived - the perpendicular velocity, as it is closely related to the Doppler shift ($v_\perp(\rho,t)=2\pi f_d/k_\perp$). In a broad sense, $v_\perp$ can be considered as the sum of $v_{E\times B}$ and $v_{\phi}$ when $v_{\phi}$ is negligibly small compared to $v_{E\times B}$ at the plasma edge. Subsequently, $v_{E\times B}$ is associated with the $E\times B$ shearing rate:

\begin{equation}
	\gamma_{E\times B}=(r/q)d(E_r/B_pR)/dr\sim \frac{\partial v_{E\times B}}{\partial r}
\end{equation}
where $q$ represents the safety factor, and $B_p$ signifies the poloidal magnetic field. As is well-established, turbulence suppression takes place when the $E \times B$ shearing rate surpasses the linear growth rate of the underlying instability, and conversely\cite{schirmer2006radial}. Given that the measured velocities are primarily influenced by $E \times B$ shear, all other factors being constant, one can estimate the evolution of $\gamma_{E\times B}$.

A high-density scenario ($n_e(0) \simeq 1.83 \sim 2.02 \times 10^{19}/m^3, I_p \simeq 160 kA$) was investigated while maintaining consistent puff intensity. Using data from the DBS reflectometer, electron temperature fluctuations in the high-density scenario were investigated and are presented in figure \ref{fig7}. In figure \ref{fig7}(a), the profile for this scenario is displayed, while figure \ref{fig7}(b) shows the radial coherent length. The $\tilde T_e/T_e$ was also integrated with 20-200 kHz coherencies to obtain the normalized coherence (20-200 kHz). The legend "w/o nonlocal, puff" indicates the period of $T_e$ drop after puffing, while "w/o puff" corresponds to the steady-state phase. In contrast to figure \ref{fig5}, the $\tilde{T_e}/T_e$ and coherent length exhibit smaller increases during the non-steady state. Importantly, the coherence entirely dissipates at $\rho\sim 0.65$. This suggests that edge puffing has minimal influence on the core in the context of relatively higher-density plasma. For the central $\tilde{T_e}/T_e$, the fluctuation amplitudes remain below the noise level and cannot be reliably extracted.

\begin{figure*}[!htbp]
	\centering
	\includegraphics[scale=0.50]{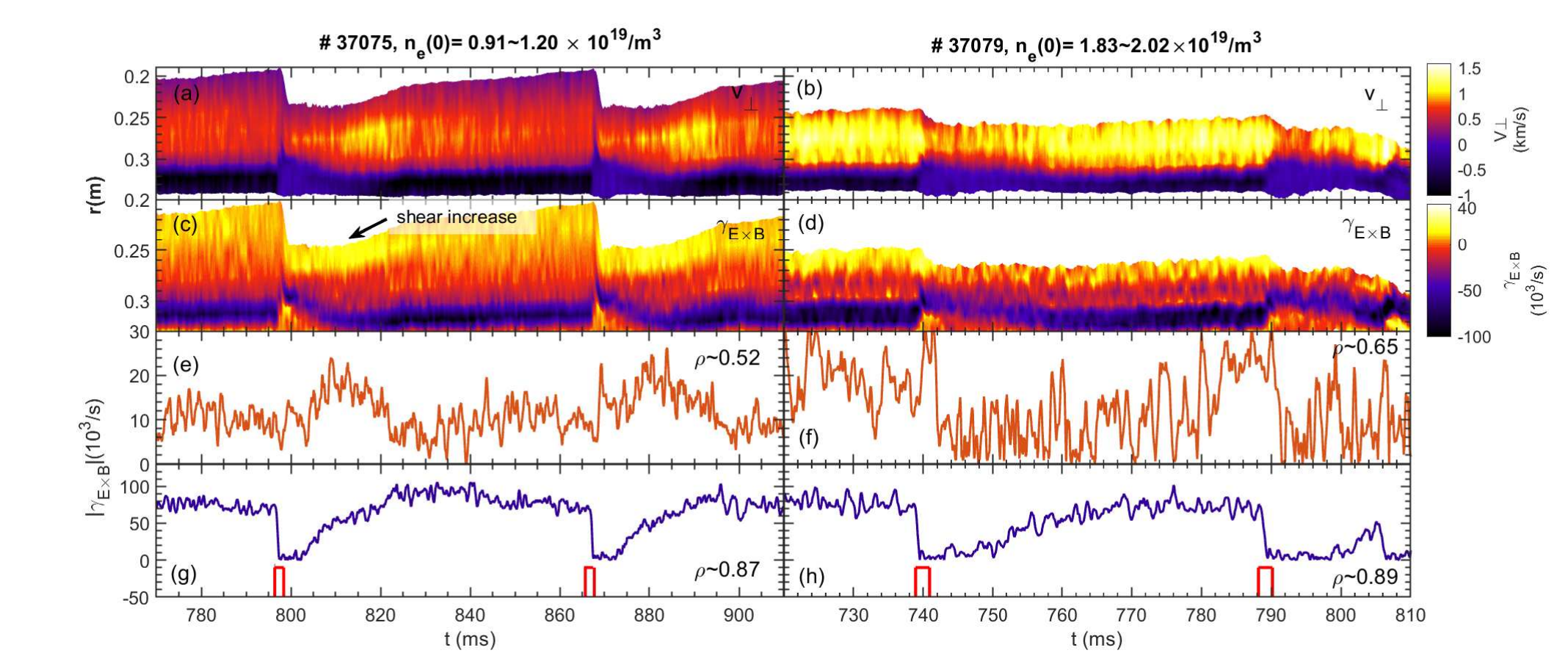}
	\caption{\label{fig6} Spatiotemporal plots of (a-b)Perpendicular velocities  $v_{\perp}$, (c-d)$E\times B$ shear rates $ \gamma_{E\times B}$ and (e-h) absolute values of $\gamma_{E\times B}$ with core $T_e$ increase (nonlocal phenomenon) in the low density  ($n_e(0) \simeq 0.91\sim 1.20 \times10^{19}/m^3$) scenario and without core $T_e$ increase in the high density ($n_e(0) \simeq 1.83\sim2.02 \times10^{19}/m^3$) scenario. The plus or minor value of $v_\perp$ represents the ion or electron diamagnetic drift direction, respectively. The shearing rates increase abruptly inside inverse region in the non-local scenario.}
\end{figure*}

Furthermore, Figure \ref{fig6} displays the $E\times B$ shearing rates ($\gamma_{E\times B}$) for this particular scenario (shot 37079) alongside data from shot 37075. The polarity of $v_\perp$ signifies ion or electron diamagnetic drift direction, respectively. It's evident that the $E \times B$ flow experiences transient variations after the puffs in both scenarios. In shot 37079, $v_\perp$ at all detected radii decelerates, causing shear rates to immediately decline with an inward propagation damping effect.
Conversely, in the low-density scenario, the reduction in $\gamma_{E\times B}$ persists in the peripheral region. However, the propagation dynamics shift within $\rho = 0.52$. Instead, $v_\perp$ and $\gamma_{E\times B}$ show rapid increases after the puffs, a phenomenon distinct from the high-density scenario. Notably, the point at which shear rates begin to change aligns closely with the $T_e$ inverse radius. In both scenarios, the trends in shear rates in the central region are consistent with their impact on $\tilde T_e/T_e(t)$.

These findings align closely with the critical gradient model described in previous works\cite{Wangzhanhui2011}. To elucidate further, considering the radial dimension, it can be expressed as follows:

\begin{align}
\frac{\partial I}{\partial \hat{t}}-\frac{\partial}{\partial \hat{\rho}}(\hat{\chi}_{turb} \frac{\partial I}{\partial \hat{\rho}})= \hat{\gamma} I- \hat{\beta}I^{2},\\
\frac{\partial \langle \hat{T} \rangle}{\partial \hat{t}}-\frac{\partial}{\partial \hat{\rho}}[(\hat{\chi}_{turb}+\hat{\chi}_{neo})\frac{\partial \langle \hat{T} \rangle}{\partial \hat{\rho}}]=\hat{S},
\end{align}
where the normalized quantities are denoted with a superscript $\hat{}$ and the normalized process was described in \cite{Wangzhanhui2011}. Here, $I$ is the turbulence intensity, $\beta$ is local nonlinear damping rate, $\hat{\chi}_{turb} = \hat{\chi}_0 I /(1+\hat{A}^2/\hat{\kappa}_c^2)$ is the normalized turbulent thermal diffusivity, $\hat\chi_{neo}$ is the normalized neoclassical thermal diffusivity , $\hat{S}$ is the normalized heating source and $\hat{\gamma}=\hat{\gamma}_0\hat{R}(\hat{A}-\hat{A}_{crit})\Theta(\hat{A}-\hat{A}_{crit})/(1+\hat{A}^2/\hat{\kappa}_c^2)$ is the local growth rate of turbulent intensity with a Heviside function $\Theta$. $\hat{A}=|\partial \langle \hat{T} \rangle|/\partial\hat{\rho}$ is the magnitude of the local mean temperature (ion or electron) gradient. The electric field shear is related to local parameter with $\hat{\kappa}_c^2$. Notably, turbulence spreading exhibits a velocity of $\hat{v}f\sim \sqrt{\hat{\gamma}\hat{\chi}_{turb}}$, which is comparable to the Fisher front speed\cite{FISHER,PhysRevLett.118.185002}. It has been demonstrated in simulations that with strong shearing (smaller $\hat{\kappa}_c$) and weak shearing, a bifurcation of $T_e$ can be achieved\cite{Wangzhanhui2011}.

\section{Conclusion and Discussion}
\label{sec:4}

This study sheds light on the intricate dynamics of turbulence, which play a pivotal role in non-local transport effects in tokamak plasmas \cite{li_cpb,li_cpb1,jiao}. Leveraging advanced core detection and measurement techniques, our analyses unveil the behavior of non-local transport in low-density plasmas, intricately linked with the broadband (20-200 kHz) electron temperature fluctuations. Both our findings in the frequency domain through cross-coherence analysis and the time-domain $\tilde{T}_e/T_e(t)$ demonstrate that transient changes are intimately associated with turbulent dynamics. During non-locality phases, the $\tilde{T}_e/T_e$ temporal profiles exhibit marked differences between the central and edge regions. Consequently, we challenge the conventional understanding of local $a/L_{T_e}$ driving mechanisms. Furthermore, our observations reveal the rapid formation of long radial coherent fluctuations, with coherence lengths comparable to the plasma size (approximately 0.5$a$). This phenomenon arises immediately after pulse injection, effectively breaking down local transport relations. In essence, transient transport can be initiated by distant effects.

Simultaneously investigating plasma rotation, we identify a partial alignment between $E\times B$ flow shear and the transient fluctuation changes, without accounting for instability growth rates. However, in a comparative high-density scenario, our observations manifest stark differences, with no clear indications of core-edge transient interactions. When the non-local conditions are met, particularly in scenarios characterized by lower density and larger plasma current, we observe a simultaneous decrease in flow shearing rates and rapid turbulence growth in the same region. This observation suggests that flow shearing rates could significantly contribute to the formation of long radial coherent structures due to their interplay with turbulence. Our direct observations show an increase in core shear rates within 1-2 ms, which corresponds to transient changes in $\tilde{T}_e/T_e$. Notably, this propagation speed of amplitude change is on the order of $10^2$ m/s, far exceeding the device's global confinement time of 30 ms. Considering that the inward coherent length extends to approximately 8.8 cm with the reference at $\rho = 0.80$, it encompasses a significant portion of the plasma's minor radius.

In addition, these observations are contingent on diagnostic compatibility, which can be further optimized. For instance, the measured $\tilde{T}_e/T_e$ profile exhibits a steep decay, reaching noise levels at relatively peripheral radii. This limitation could be mitigated by employing a smaller beam size or extending the Rayleigh length of the diagnostic sight. Additionally, the $v_\perp$ of the flow will be detected in the central core using a higher radio frequency of the reflectometer. The observed fluctuation bifurcation hints at the potential formation of a barrier-like structure, akin to a transient internal transport barrier (ITB) \cite{xu}, in the inverse region. Evidence of interactions between shear profiles and structures can also be gleaned from comparative experiments, where the location of the barrier (sudden change) varies. Importantly, these results are corroborated by their disappearance or significant weakening in scenarios without non-locality.


In general, local plasma parameters play a pivotal role in driving instabilities, such as pressure gradients, ion and electron temperature gradients. Even in cases of non-diffusive momentum transport, which falls under the umbrella of locality since these non-diffusive terms are determined by local parameters, non-local transport exhibits a clear departure from local closure. Non-local transport encompasses the influence of parameters located far from where the transport occurs, challenging the notion that electrostatic turbulence alone dictates heat flux. Instead, it is primarily determined by the coupling between turbulence and electric fluctuations. Nonetheless, the presence of shear and long radial coherence in electron temperature fluctuations provides valuable insights. This study does not delve into other processes, such as nonlinear multi-scale turbulent interactions, turbulence spreading, and density-temperature bicoherence. However, it's worth noting that in the theory of self-organized criticality, enhanced avalanches can occur when a long radial correlation is present \cite{pan2015evidence}. This could potentially serve as a mechanism for transient flow variations. To conduct a comprehensive investigation, future experiments will explore electric field, density, and temperature channels. 

{\it Acknowledgement}
The authors thank to Prof. Inagaki for useful analytic discussions in APTWG meeting. This work is partly supported by National Key R\&D Program of China (Grant Nos. 2017YFE0301203, 2017YFE0301106), Sichuan Science and Technology Program (Grant No. 2018RZ0123, 2021YFSY0044), National Natural Science Foundation of China (Grant Nos. 12175055), and Shenzhen Municipal Collaborative Innovation Technology
Program-International Science and Technology Cooperation Project (GJHZ20220913142609017).

\end{document}